\documentclass[10pt, conference, compsocconf]{IEEEtran}
\ifCLASSINFOpdf
 \usepackage[pdftex]{graphicx}
\else
\fi
\usepackage{array}
\newcolumntype{L}[1]{>{\raggedright\let\newline\\\arraybackslash\hspace{0pt}}m{#1}}
\newcolumntype{C}[1]{>{\centering\let\newline\\\arraybackslash\hspace{0pt}}m{#1}}
\newcolumntype{R}[1]{>{\raggedleft\let\newline\\\arraybackslash\hspace{0pt}}m{#1}}
\usepackage{amsmath} 
\usepackage[hyphens]{url}
\usepackage[T1]{fontenc}
\usepackage[hidelinks,bookmarks=false]{hyperref}
\newcommand{\reffig}[1]{Figure~\ref{#1}}

\hypersetup{pdfstartview={XYZ null null 1.00}}
\IEEEoverridecommandlockouts
\begin{document}
%
\title{When Edge Computing Meets Compact Data Structures
\thanks{This research is supported in part by Chilean National Research and Development Agency (ANID, Chile) under Grant FONDECYT Iniciaci{\'o}n 11180905.}
}

\author{\IEEEauthorblockN{Zheng Li}
\IEEEauthorblockA{\textit{Department of Computer Science} \\
\textit{University of Concepci{\'o}n}\\
Concepci{\'o}n, Chile \\
ORCID: 0000-0002-9704-7651}
\and
\IEEEauthorblockN{Diego Seco}
\IEEEauthorblockA{\textit{Department of Computer Science} \\
\textit{University of Concepci{\'o}n}\\
Concepci{\'o}n, Chile \\
Email: dseco@udec.cl}
\and
\IEEEauthorblockN{Jos{\'e} Fuentes}
\IEEEauthorblockA{\textit{Department of Computer Science} \\
\textit{University of Concepci{\'o}n}\\
Concepci{\'o}n, Chile \\
Email: jfuentess@udec.cl}
}

\maketitle

\begin{abstract}
Edge computing enables data processing and storage closer to where the data are created. Given the largely distributed compute environment and the significantly dispersed data distribution, there are increasing demands of data sharing and collaborative processing on the edge. Since data shuffling can dominate the overall execution time of collaborative processing jobs, considering the limited power supply and bandwidth resource in edge environments, it is crucial and valuable to reduce the communication overhead across edge devices. Compared with data compression, compact data structures (CDS) seem to be more suitable in this case, for the capability of allowing data to be queried, navigated, and manipulated directly in a compact form. However, the relevant work about applying CDS to edge computing generally focuses on the intuitive benefit from reduced data size, while few discussions about the challenges are given, not to mention empirical investigations into real-world edge use cases. This research highlights the challenges, opportunities, and potential scenarios of CDS implementation in edge computing. Driven by the use case of shuffling-intensive data analytics, we proposed a three-layer architecture for CDS-aided data processing and particularly studied the feasibility and efficiency of the CDS layer. We expect this research to foster conjoint research efforts on CDS-aided edge data analytics and to make wider practical impacts.
\end{abstract}


%
\IEEEpeerreviewmaketitle

\section{Introduction}
Edge computing is a modern distributed computing paradigm that brings data processing and storage closer to the network edge where the data are generated. Given the pervasive and heterogeneous edge and user devices, both the deployment environment and the runtime topology can be more distributed than ever for edge applications \cite{Cooke_2020}. Correspondingly, it has been envisioned that the demands of data sharing and collaborative processing will dramatically increase on the edge \cite{Zhang_Zhang_2016}. For example, in the field of edge intelligence, to label cross-region data for training AI models, the preprocessing of distributed raw data will be needed by employing data-parallel frameworks like MapReduce \cite{Jin_Jia_2021}.

It should be noted that the data-parallel processing can be data-shuffling intensive. In cluster computing, the data shuffling between computation stages accounts for more than half of the completion times of MapReduce and Dryad jobs \cite{Chowdhury_2011}. In Amazon cloud, data communication at the shuffling stage can even take up to 70\% of the overall execution time of self-join applications \cite{Zhang_2013}. Considering the limited power supply for wireless equipment and the limited  bandwidth resource in the edge environments, such excessive communication loads will inevitably lead to threats not only to the time critical requirement of edge computing but also to the battery life of edge devices. Therefore, it is crucial and valuable to investigate practical approaches to reducing the communication overhead in edge data analytics.

A straightforward strategy of reducing the amount of data that need to be transmitted over a network is to employ data compression techniques \cite{Gia_2019}. However, this strategy is mainly suitable for one-off data transmission scenarios, as data decompression is generally required to utilize the compressed data, no matter in case of lossy compression or lossless compression. 
When it comes to shuffling-intensive data analytics, the extremely frequent data compression and decompression will explosively increase the computation workloads, which will not pay off the reduced communication overhead.  

Compared with data compression, the techniques of compact data structures (CDS) can not only maintain data with less space, but they also enable the data to be queried, navigated, and manipulated directly in their compact form, i.e.~without being decompressed \cite{Navarro_2016}. Such a decompression-free feature makes CDS particularly promising to fit in the scenario of space-sensitive data analytics. As a matter of fact, CDS techniques have been successfully applied to many big data areas (e.g., bioinformatics \cite{Makinen_2015}). However, there exist various challenges in applying CDS to edge computing. For example, despite the reduced data size, the construction of CDS is still space hungry, which may not be affordable for resource-limited devices. Although the edge computing community has noticed and recognized the benefits of CDS, the current discussions about employing CDS techniques are superficial and ad hoc \cite{Buddhika_2021}, not to mention the lack of empirical experience reports and case studies.     

To help gain deeper understanding of, and obtain the first-hand experience in, applying CDS to edge computing, we firstly brainstormed a set of challenges and opportunities in practice, and then adopted Arduino Y{\'U}N Rev2 as a representative edge device to investigate CDS-aided data analytics implementations. This paper reports our current work with a twofold contribution.  

\begin{itemize}
\item For researchers, our discussions about the challenges and opportunities of applying CDS to edge computing can act as a bridge between these two communities, to foster conjoint research efforts on the topic of CDS-aided edge data analytics. 
\item For practitioners, our empirical study demonstrates the feasibility and potential efficiency of implementing CDS-aided edge data analytics. The three-layer architecture proposed in our study can be adapted to more use cases on the edge. 
\end{itemize}

The remainder of this paper is organized as follows. Section \ref{sec:challenges} summarizes our brainstormed challenges, opportunities, and two main implementation scenarios of applying CDS to edge computing. Section \ref{sec:experiment} describes our empirical study that mainly focuses on the CDS layer at this current stage. Conclusions are drawn in Section \ref{sec:conclusion} together with our future work plans.

\section{Challenges and Opportunities of Applying CDS to Edge Computing}
\label{sec:challenges}

\subsection{Major Challenges}
Although the benefit from compact while still operable data is intuitively straightforward, applying CDS to edge computing may have various challenges in practice.

\begin{itemize}
\item \textbf{The construction of CDS is space hungry.} As a cost of enabling data operations in
the reduced space, extra runtime space is needed for constructing CDS. Despite active research efforts in the last few years,
many CDS require a large construction space (that may be tremendously larger than the constructed results). Given those IoT devices with limited storage capabilities (e.g., the embedded RAM is only 4KB for current taxi-mounted GPS devices), the construction of CDS will be one of the top challenges on the edge.

\item \textbf{The dynamism of CDS construction increases the computational complexity.} To address the space limit, an alternative strategy is to dynamically construct CDS via update operations for chunk-by-chunk datasets. Take our empirical study as an example (cf.~Section \ref{subsec:CDStechniques}), instead of constructing and maintaining a full dictionary to encode data, we can let edge devices build up and continuously update a dictionary subset for data encoding. Unfortunately, this strategy will inevitably increase the computing overhead on the edge. Moreover, although CDS aims to eliminate as much redundancy as possible, such a CDS dynamism requires redundant information to allow update operations and to make dynamic CDS constructions compatible with each other. In fact, it has been revealed that the dynamic versions of typical CDS techniques do not achieve more remarkable performance than their static versions \cite{Navarro_2016}. 

\item \textbf{Distributed CDS constructions may incur CDS inconsistency issues.} Given the largely dispersed data distribution on the edge, CDS constructions should also be arranged on distributed edge devices, in order collaboratively to satisfy the needs of compact data consumption. When heterogeneous data and dynamism are involved in the distributed CDS constructions, the update operations can make the constructed CDS inconsistent on different edge devices. To our best knowledge, this challenge has not been well studied even in the CDS community. A possible solution is to develop a CDS transmission mechanism. One device can transmit its CDS (or a part of it via split/filter operation) to another device, or even to a centralized CDS server. The receivers will be able to merge different CDS versions and synchronize the others with a consolidated version.  
\end{itemize}

\subsection{Promising Opportunities}
As long as CDS is implementable on suitable edge devices (by tolerating, bypassing, relieving or addressing the aforementioned challenges), we argue that it can bring at least two main benefits to edge computing, as explained below.

\begin{itemize}
\item \textbf{Reducing the communication overhead in shuffling-intensive edge data analytics.} Data shuffling is a crucial component in collaborative data analytics across multiple compute nodes (e.g., MapReduce). As mentioned previously, data communication at the shuffling stage can dominate the overall execution time of distributed computing jobs \cite{Chowdhury_2011,Zhang_2013}. However, such a communication overhead can result in violations of the low-latency requirements of edge computing. In fact, it has been identified that data shuffling has become a bottleneck for edge data analytics \cite{Zhao_Wang_2019}. By applying CDS, the largely reduced data size will significantly reduce the communication overhead of data shuffling. It should be noted that the data compression techniques could not be applicable in this case, as data shuffling always comes with intermediate data processing, while the compressed data would not be able to be processed directly.  
 
\item \textbf{Extending the battery life of edge devices.} Unless using wired power supply, the usage of edge devices is mainly constrained by their battery capacities, while there is still a tremendous gap between battery technologies and power requirements due to the inherent complexity in the relevant interdisciplinary topics (e.g., thermodynamics and fluid mechanics) \cite{Li_Pradena_2021}. Meanwhile, given the massively growing IoT and the increasing popularity of over-the-top mobile applications (e.g., instant messaging), signaling energy consumption has become a major concern on the edge \cite{Chan_Li_2015}. Therefore, the data transmission among edge devices should take into account not only offering high throughput and low latency, ``but also conserving precious battery energy to prolong operational lifetime'' \cite{Fang_Li_2014}. Considering the benefit of maintaining the same data usability after reducing the data size, CDS-aided data transmission can be a promising strategy to extend the battery life of relevant edge devices.
\end{itemize}

\subsection{Potential Implementation Scenarios}
When it comes to practically implementing CDS techniques on edge devices, we distinguish between two main implementation scenarios.

\begin{itemize}
\item \textbf{Static CDS based on prior knowledge.} In this case, the CDS can be constructed in advance based on pre-known regular patterns or domain-specific knowledge, and correspondingly the compact counterpart of the original data form will have been predefined. At runtime, the constructed CDS can act as a static transformer to convert original data to their compact versions seamlessly. This scenario is exemplified by a pre-established lookup table in our empirical study (cf.~Section \ref{subsec:CDStechniques}).
\item \textbf{Dynamic CDS based on runtime knowledge.} In this case, the prerequisite information for constructing CDS is unavailable until the data to be converted are received at runtime, and frequent CDS reconstructions may be needed due to possibly continuous information updates (e.g., the price distribution of Amazon's spot service). It is clear that such a scenario would also suffer from the dynamism challenge of CDS construction, although we could employ buffers to facilitate the storage of new data and the reconstruction of CDS. In practice, a possible workaround is to accumulate and take advantage of the posterior knowledge (e.g., the price distribution may exist only in a limited range), to reduce the frequency of CDS reconstruction.
\end{itemize}


\section{An Empirical Investigation}
\label{sec:experiment}
\begin{figure}[!t]
\centerline{\includegraphics[width=7.6cm,trim=60 610 355 60,clip]{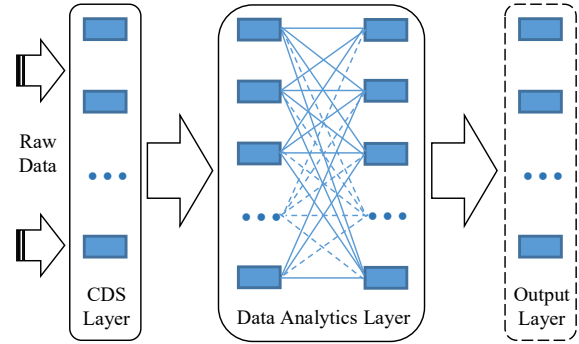}}
\caption{The three-layer architecture for CDS-aided edge data analytics. (\textit{Each blue rectangular indicates an edge device.})}
\label{fig:scenario}
\end{figure}

Instead of arguing specific CDS techniques (e.g., bitvectors, wavelet trees, $k$-page graph, etc.) \cite{Navarro_2016}, this research aims to bring us empirical experience in applying CDS to edge computing in a generic sense. Therefore, we decided to firstly propose a generic edge computing scenario that involves CDS, in order to better drive our experimental investigation. In this scenario, we design  a three-layer architecture for conducting suitable data analytic tasks at the edge side, as illustrated in \reffig{fig:scenario}. 

In detail, while receiving raw data, edge devices at the first layer choose suitable CDS techniques to transform the raw data, and continuously feed the second layer with data in a compact form. A group of second-layer devices perform data analytics in a collaborative and distributed manner, without even being aware of the original data form. Depending on the requirement, the data analytical results can be output directly or be switched back to the plain form. Thus, the third layer is optional in case the reverse data transformation is needed. 

As for the data analytic tasks, we refer to our previous work on characterizing Amazon's spot service pricing \cite{Li_Li_Li_2019}. Naturally, the whole system will need to be re-implemented and deployed in the edge environment. More importantly, the workflow will need to be adapted to the three-layer architecture designed in this research, and using compact price records instead of original price history to realize the characterization. Driven by the aforementioned concerns about edge resource constraints when applying CDS, we only focus on the first layer in this paper.

\subsection{Selecting Suitable CDS Techniques}
\label{subsec:CDStechniques}
Given the spot service prices downloadable from Amazon (each record includes the tag, price, timestamp, instance type, operating system, and AWS zone, as exemplified in \reffig{fig:original}), we see four strategies to make the original data compact, such as:

\begin{figure*}[!t]
\centerline{\includegraphics[width=17.25cm,trim=45 600 45 60,clip]{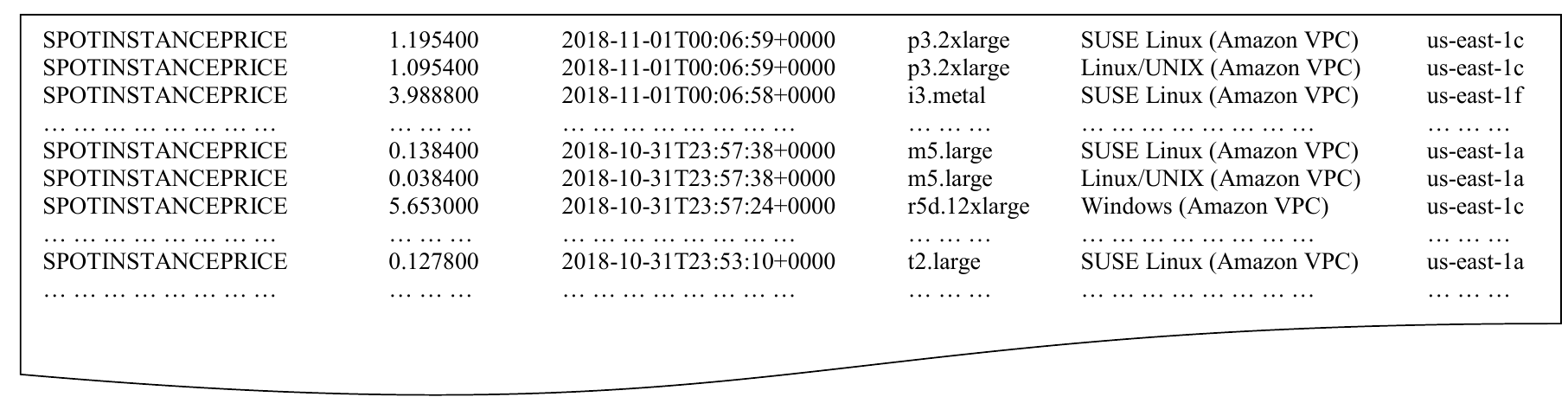}}
\caption{Original spot service price records downloadable from Amazon. The columns from left to right are: Tag, Price, Timestamp, Instance Type, Operating System, AWS Zone.}
\label{fig:original}
\end{figure*}

\begin{itemize}
\item \textbf{Filtering out redundant data components.} In the context of spot service, the price record tag ``SPOTINSTANCEPRICE'' never changes, and the last two digits of the prices are always 0. Thus, we can directly ignore them when reading the data. Since Amazon adopts UTC by default \cite{Amazon_2021}, the UTC offset part (i.e.~``+0000'') of the timestamp can also be removed. 
\item \textbf{Delta encoding data via differential compression.} The differential strategy is widely shared between compression techniques and CDS techniques. In our case study, this strategy is particularly suitable to reduce the size of continuous timestamps. By setting a base timestamp, the following timestamps can be represented as their differences against the base one. In practice, multiple base timestamps with suitable intervals can be used to control the maximum time difference. 
\item \textbf{Using prior knowledge to encode data.} Amazon defines its unique service (EC2) instances via a composite key that is made up of instance type, operating system and AWS zone. There three attributes all have fixed amounts of values (i.e.~402 types, eight systems, and 56 zones respectively) \cite{Amazon_2021}. Benefiting from this prior knowledge, we can encode those unique service instances (e.g., into Huffman codes), and then store the codes in bitvectors.  
\item \textbf{Using posterior knowledge to encode data.} Although the spot pricing mechanism delivers dynamic prices at runtime, each service instance's price varies only in a limited range. Considering that the pricing characterization will deliver a price distribution (as demonstrated in \cite{Li_Li_Li_2019}), after data analytics, we can utilize the price distribution to further encode the price numbers and in turn improve the CDS efficiency in future characterization work.
\end{itemize}

The first two strategies are straightforward to implement. Particularly, to make the demonstration reader-friendly in this paper, we still keep the base timestamps in the human-readable format. In contrast, since the data analytics layer is out of our current research scope, the fourth strategy is not implemented in this study.

When implementing the third strategy, for the purpose of conciseness and for the ease of replication, we decided to use sequential numbers instead of binary codes to represent and integrate the available information of instance types, operating systems and AWS zones. In fact, there is always a CDS trade-off between ``good space performance of bitwise codes and the good time performance of bytewise codes'' \cite{Navarro_2016}. Eventually, we established a lookup table for Arduino Y{\'U}N Rev2 to encode incoming data.

\subsection{Setting Up the Testbed}
\begin{figure}[!t]
\centerline{\includegraphics[width=7.2cm,trim=60 615 330 60,clip]{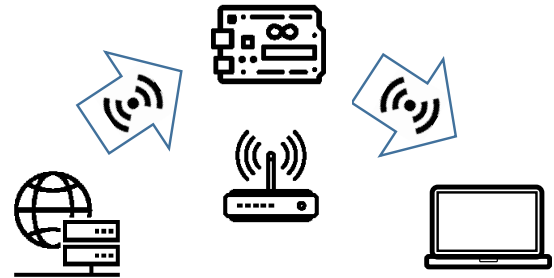}}
\caption{The testbed used in our experimental investigation.}
\label{fig:testbed}
\end{figure}

By sticking to the CDS layer, our experimental logic is focused on an edge device that can receive plain spot prices and transmit out their compact version. Accordingly, we follow the client-server model to architect and build up the testbed, as shown in \reffig{fig:testbed} and explained below.

\begin{itemize}
\item \textbf{A Wireless Router:} We work in a stable WiFi environment via a TP-Link Realtek RTL8821CE 802.11ac PCIe Adapter that transmits and receives data on the frequency 2.4 GHz and at the speed 150 Mbps. 
\item \textbf{An Arduino Y{\'U}N Rev2:} To represent typical edge device features, we choose an Arduino Y{\'U}N Rev2 that has built-in Ethernet and WiFi support. In particular, we intentionally utilize its limited storage and compute capacities by sticking to its native sketch side for all the experiments. Correspondingly, the aforementioned CDS strategies are implemented as Arduino sketch program.
\item \textbf{An HTTP Server:} Since the sketch side of Arduino Y{\'U}N Rev2 does not have enough power to handle HTTPS connections, we set up a simple HTTP server on a Windows 10 desktop to simulate the data source, by hosting various sizes of Amazon's spot service pricing history files.
\item \textbf{A Client:} To facilitate our observations and measurements, 
we issue data requests through a client Python program deployed on a Windows 10 laptop. Each client request triggers a cascade Arduino request to the HTTP server, while the eventual response will be the CDS processing results instead of the raw data.  
\end{itemize}

\subsection{Experimental Design and Implementation}
\label{subsec:design}
Since the resource limit is the major concern for applying CDS on edge devices (cf.~Section \ref{sec:challenges}),
we have prepared a set of data files ranging from 5 to 1000 price records to try gradually squeezing the capacity of Arduino Y{\'U}N Rev2. For each data file hosted on the HTTP server, we issue multiple requests and measure the average wall-clock latency from the client. In particular, the data processing latencies on Arduino Y{\'U}N Rev2 are measured via the sketch code \texttt{millis();} and also returned to the client. It should be noted that both the cascade request time and the response time have been excluded from the measurement of data processing latency.

The data processing here is to apply the CDS strategies to the original price records, as discussed in Section \ref{subsec:CDStechniques}. 
To improve the replicability of this research, we share the source files online\footnote{\url{https://www.doi.org/10.5281/zenodo.5149326}} and describe the main steps as follows.

\begin{enumerate}
\item Launch the HTTP server, and keep note of the server's IP address and port number. 
\item Update the \texttt{read\_file()} function in the sketch code with the noted IP address and port number.
\item Power on Arduino Y{\'U}N Rev2 and upload the updated sketch code to it. 
\item Inject the pre-established lookup table as key-value pairs to Arduino Y{\'U}N Rev2. 
\item Issue client requests to obtain spot price data and latency measurement results.
\end{enumerate}

\begin{figure}[!t]
\centerline{\includegraphics[width=5.95cm,trim=45 615 395 50,clip]{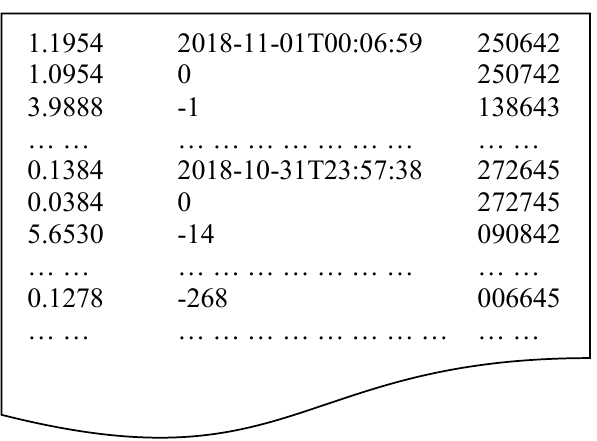}}
\caption{Human-readable compact form of the sample data after the CDS conversion. (\textit{In production, the bitwise machine-readable form can be more compact.})}
\label{fig:less}
\end{figure}

\subsection{Experimental Results and Analyses}
After going through the CDS conversion, the original price records demonstrated in \reffig{fig:original} will become as compact as shown in \reffig{fig:less}. Due to the known timeout issue\footnote{\url{https://github.com/espressif/arduino-esp32/issues/1433}}, in our tests, Arduino Y{\'U}N Rev2 can barely cope with data beyond 650 price records for one request. Therefore, we only conducted experiments with data amounts up to 650 records. The experimental results corresponding to different data sizes are exemplified in Table \ref{table_results}.

\begin{table}[!t]
\renewcommand{\arraystretch}{1.3}
\caption{Selected Latency Measurement Results with respect to Different Data Sizes (Record Amounts)}
\label{table_results}
\centering
\begin{tabular}{|C{1cm}|C{1.3cm}|C{1.3cm}|C{1.3cm}|C{1.3cm}|}
\hline
\textbf{Record Amount} & \textbf{CDS Wall-clock Latency} & \textbf{CDS Processing Latency} & \textbf{Non-CDS Wall-clock Latency} &\textbf{Non-CDS Processing Latency}\\
\hline
12 & 1.852s  & 0.735s & 1.921s & 0.539s \\
\hline
25 & 2.935s & 1.765s & 2.992s & 1.012s\\
\hline
50 & 5.039s & 3.533s & 4.902s & 1.935s\\
\hline
100 & 9.148s & 6.950s & 8.672s & 3.830s\\
\hline
200 & 17.258s & 13.791s & 16.633s & 7.713s\\
\hline
400 & 33.588s & 27.199s & 31.837s & 14.868s\\
\hline
600 & 50.765s & 41.630s & 48.487s & 23.314s \\
\hline
\end{tabular}
\end{table}

Note that in addition to the predesigned experiments (cf.~Section \ref{subsec:design}), we also measured baseline latencies, by treating Arduino Y{\'U}N Rev2 as a relay. To avoid relaying data byte by byte, we let Arduino Y{\'U}N Rev2 cache the incoming bytes and transmit individual records to the client. As such, we define the caching overhead as the non-CDS processing latency and distinguish it from the non-CDS wall-clock latency. 

\begin{figure}[!t]
\centerline{\includegraphics[width=8cm,trim=0 7 0 0,clip]{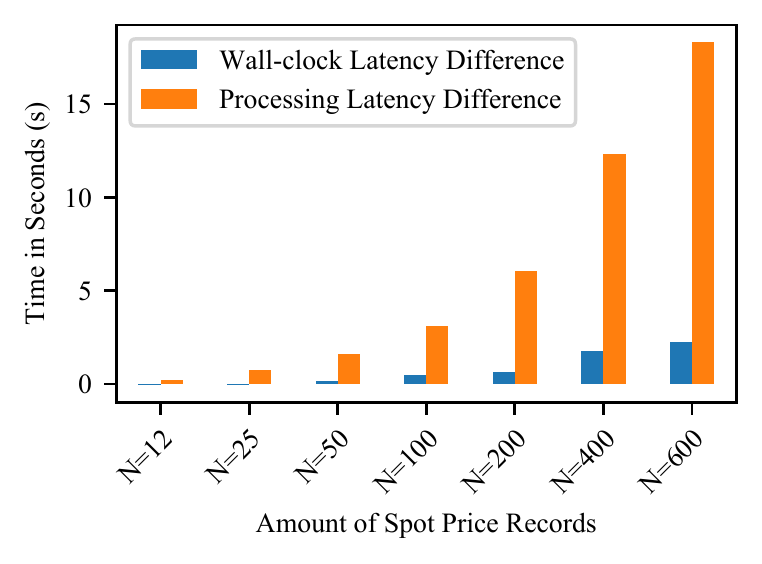}}
\caption{Latency difference between the CDS and the non-CDS scenarios.}
\label{fig:difference}
\end{figure}

From the numerical results in Table \ref{table_results}, it is unsurprising to see that the CDS processing always takes more time than its corresponding baseline. According to the No-Free-Lunch theorem \cite{Rabhi_Bandara_2018}, the expected communicational benefit in the data analytics layer inevitably requires the computational cost in the CDS layer. However, when the data size is small (e.g., the record amount is 12 or 25), CDS seems to be able to bring extra benefits for the data communication, i.e.~the reduced data size also lowers the wall-clock latency. To facilitate the observation, we further visualize the latency difference between the CDS and the non-CDS scenarios, as shown in \reffig{fig:difference}.

\begin{footnotesize}
\begin{equation}
\label{eq_ratio}
\frac{\text{Computation}}{\text{Communication}} = \frac{\text{Processing Latency}}{\text{Wall-clock Latency}-\text{Processing Latency}}
\end{equation}
\end{footnotesize}

Benefiting from the visualization, it is not only clear that the latency difference keeps increasing along with the growth of data size, but that the processing latency difference increases significantly faster than the wall-clock latency difference. In other words, by referring to Eq.~(\ref{eq_ratio}), the computation-to-communication ratio will become higher and higher when the CDS workload increases. Recall that distributed systems generally favor high computation-to-communication ratio for various purposes, ranging from efficient resource utilization to improved scalability \cite{Li_Maddah-Ali_2018}. From the perspective of a whole analytic task, we should reasonably maximize the transactional workloads on CDS processing units, while ignoring the aforementioned trivial extra benefits from small data sizes.

\section{Conclusions and Future Work}
\label{sec:conclusion}
By compacting data while still allowing the data to be queried, navigated, and manipulated without decompressing them, the CDS techniques are particularly promising for collaborative data analytics on the edge. However, applying CDS to edge computing may suffer from various challenges in different implementation scenarios. We argue that conjoint research efforts from relevant communities should be made to boost the emerging area of CDS-aided edge data analytics. 

After prototyping the static CDS implementation on an Arduino Y{\'U}N Rev2, our future work will be unfolded along two directions. Firstly, we will extend this prototyping work to the data analytics layer and empirically study the impacts of the current CDS implementation. Secondly, we will start investigating dynamic CDS techniques in more complex use cases on the edge.


%
\newcommand{\BIBdecl}{\setlength{\itemsep}{1 ex}}
\bibliographystyle{IEEEtran}
{\footnotesize
\bibliography{CDSref}
}

\end{document}